\documentclass[aps,prl,twocolumn,superscriptaddress]{revtex4}
\usepackage{amssymb}
\usepackage{graphicx}
\usepackage{amsmath}
\usepackage{color}

\begin{document}

\title{Quantum Phase Transitions of the Bose-Hubbard Model inside a Cavity}

\author{Yu Chen}
\email{spaceexplorer@163.com}
\affiliation{Center for Theoretical Physics, Department of Physics, Capital Normal University, Beijing, 100048, China}
\author{Zhenhua Yu}
\email{huazhenyu2000@gmail.com}
\affiliation{Institute for Advanced Study, Tsinghua University, Beijing, 100084, China}
\author{Hui Zhai}
\email{hzhai@mail.tsinghua.edu.cn}
\affiliation{Institute for Advanced Study, Tsinghua University, Beijing, 100084, China}
\date{\today}

\begin{abstract}

The superfluid to Mott insulator transition and the superradiant transition are textbook examples for quantum phase transition and coherent quantum optics, respectively. Recent experiments in ETH and Hamburg succeeded in loading degenerate bosonic atomic gases in optical lattices inside a cavity, which enables the first experimental study of the interplay between these two transitions. In this letter we present the theoretical phase diagram for the ETH experimental setup, and determine the phase boundaries and the orders of the phase transitions between the normal superfluid phase, the superfluid with superradiant light, the normal Mott insulator and the Mott insulator with superradiant light. We find that in contrast to the second-order superradiant transition in a weakly interacting Bose condensate, strong correlations in the superfluid nearby a Mott transition can render the superradiant transition to a first order one. Our results will stimulate further experimental studies of interactions between cavity light and strongly interacting quantum matters. 

\end{abstract}

\maketitle

The field of cavity QED has focused on studying the consequences of strong couplings between the internal degrees of freedom of an atom and a single mode of a light field \cite{cavity}. One of the most famous models of this field is the Dicke model, where each atom is simply described by a two-level system and both the spatial motions of atoms and the interactions between atoms are ignored. The coherent coupling between atoms and the cavity light leads to a collective phenomenon known as `` superradiance " \cite{Dicke}. On the other hand, recent developments in ultracold atomic gases have greatly advanced our understanding of strongly interacting quantum gases, examples of which include the Hubbard model in optical lattices and the unitary Fermi gas with a large scattering length \cite{int_review}. 
Combining these two directions, the interdisciplinary field of ultracold atoms inside a cavity opens up opportunities to study new phenomena due to the interplay between strongly interacting quantum matters of atoms and their interaction with the single mode cavity light, which are expected to be much richer than those described by the Dicke model. Quite a few recent theoretical works have studied the Bose-Hubbard model (BHM) \cite{Rit05, Lew08, Simons09, Simons10, Ciuti13, Hemmerich14, Benitez15} and the unitary Fermi gas inside a cavity \cite{Chen}.

\begin{figure}[tbp]\centering
\includegraphics[width=2.5in]
{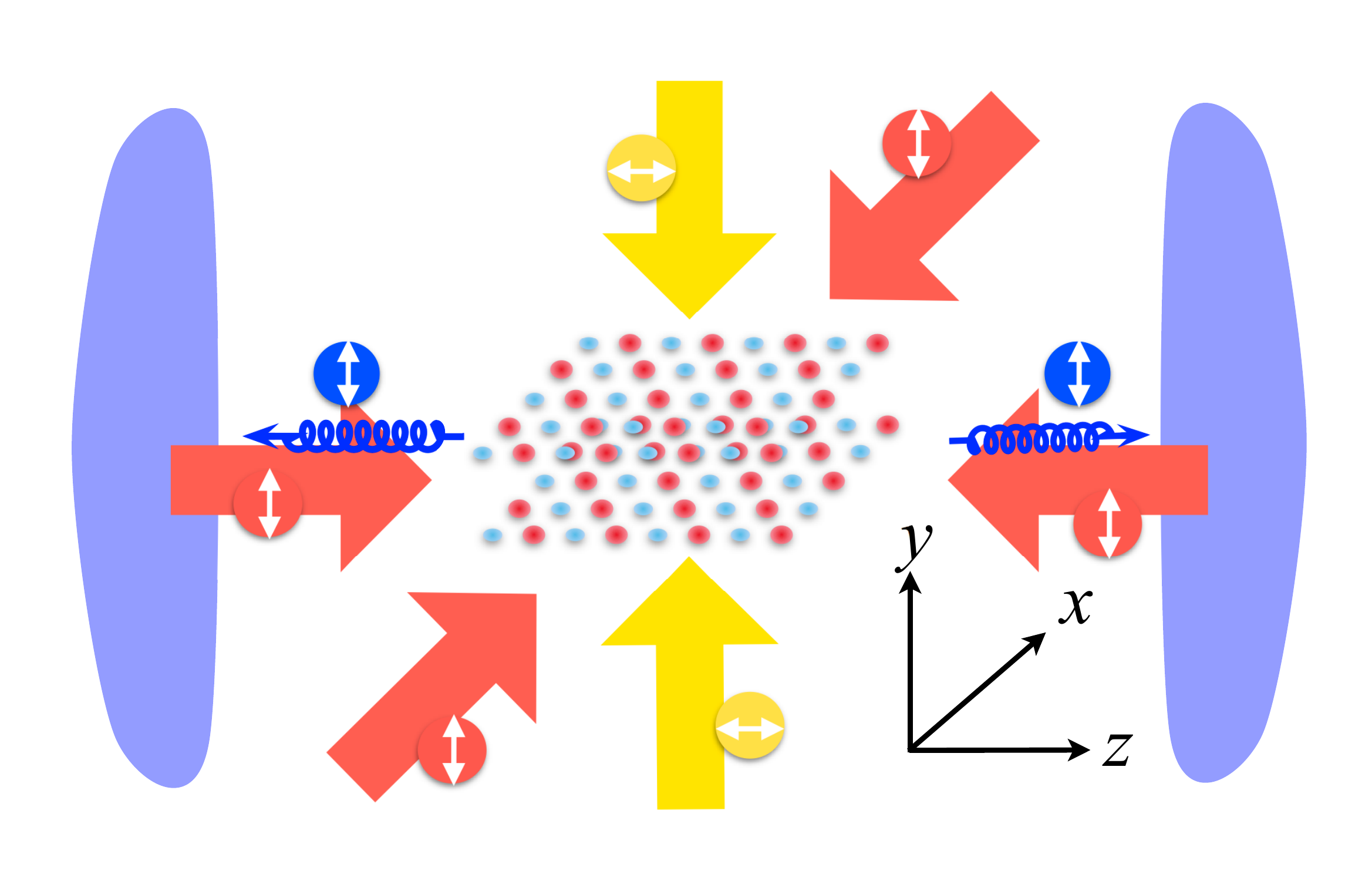}
\caption{Schematic of the ETH experiment setup. Three linearly polarised laser beams form a three-dimensional optical lattice, and a cavity is placed along the $z$-direction. \label{schematic}}
\end{figure}

Lately, both the Hamburg and the ETH groups have experimentally succeeded in loading three-dimensional degenerate bosonic atomic gases in optical lattices inside a cavity, with slight difference in their setups \cite{Hamburg,ETH}. The ETH group used a pumping beam (along the $\hat{x}$ direction) and two external lattice beams (along the $\hat{y}$ and $\hat{z}$ direction) to form a three-dimensional lattice, as schemed in Fig.~1, in which the bosonic atoms can undergo a superfluid (SF) to Mott insulator (MI) transition when the lattice depth increases. In addition, solely tuning the strength of the pumping beam can drive a superradiant transition where the expectation value of the cavity light field (along the $\hat{z}$ direction) changes from zero to finite. 
Once inside the superradiant phase, the interference between the cavity light and the pumping light gives rise to an additional check-board lattice in the $xz$ plane where a charge-density-wave (CDW) order of the bosons forms accordingly \cite{review}, and the superradiant phase transition of the cavity light is accompanied with a CDW transition of the bosons. The concurrence of the two transitions has already been observed for bosons in a cavity without lattice  \cite{Tilman}. 

Therefore, the interplay between the SF-MI transition and the superradiant transition leads to four different phases of the system: (i) the normal superfluid phase in the absence of the cavity light and the CDW order, denoted by \textit{SF}; (ii) the superfluid with the cavity light and the CDW order, denoted by \textit{CDW-SF} here and also called ``super-solid" in Ref.~\cite{ETH}; (iii) the Mott insulator without the cavity light and the CDW order, denoted by \textit{MI}; and (iv) the Mott insulator in the presence of the cavity light and the CDW order, denoted by \textit{CDW-MI} here. A rich phase diagram consisting of all these four phases has been measured and presented in the recent ETH experiment \cite{ETH}. However, none of the previous theoretical works on related subjects has addressed the same model as the ETH experimental setup, and theoretical understanding of the transitions between these phases is still lacking. This letter aims at understanding the phase diagram and the quantum phase transitions in this system.

\textit{Model and Method.} Same as the ETH experiment \cite{ETH}, here we consider a gas of bosonic atoms loaded in a three-dimensional optical lattice potential $V_\text{L}({\bf r})=V_y\cos^2(k_0y)+V_{\text{2D}}[\cos^2(k_0x)+\cos^2(k_0y)]$ created by the pumping laser and two external lattice beams inside a cavity. When the pumping laser strength $\sim \eta_0^2$ is sufficiently strong, the system is in the superradiant phase, and the single mode cavity light, denoted by the field operator $\hat a$, gives rise to an extra lattice potential $V_\text{C}({\bf r})=\eta_0\cos(k_0x)\cos(k_0z)(\hat{a}^\dag+\hat{a})+V_\text{c}\hat{a}^\dag\hat{a}\cos^2(k_0 z)$. Here, for simplicity, we consider that all light fields have the same wave-vector denoted by $k_0$, and we define the recoil energy $E_\text{r}=k_0^2/2m$ for later use. For deep enough lattice, the single-band approximation works well enough, and we obtain a BHM as
\begin{align}
\hat{\cal H}_{\text{BH}}=&\sum_{\langle{i},{j}\rangle}-t_{\langle ij\rangle} \left({\hat b}_{ i}^\dag {\hat b}_{j}^{}+\text{h.c.}\right)+\eta(\hat{a}^\dag+\hat{a})\sum_{i}(-1)^{i_x+ i_z}\hat{n}_{i}\nonumber\\
&+\frac{U}{2}\sum_{ i}\hat{n}_{ i}(\hat{n}_{i}-1)-\mu\sum_{i}\hat{n}_{ i}, \label{TB_Model}
\end{align}
where ${\hat b}_{ i}$ is the field operator for the bosonic atoms on the $i$th lattice site, $\hat{n}_{ i}={\hat b}_{ i}^\dagger {\hat b}_{ i}$, $\mu$ is the chemical potential, $\langle i j\rangle$ denotes all nearest neighboring sites, and $t_{\langle ij\rangle}$ include $t_x$, $t_y$ and $t_z$ for hopping along three different directions. 
Since we are interested in the vicinity of the SF-MI transition and the superradiant transition, where the external optical lattice is already quite deep while the cavity light is rather weak, we neglect the dependence of $t_{\langle ij\rangle}$ and $U$ on the cavity field, and we have numerically verified that the correction to this approximation is only few percent in the regime of the phase diagram presented below \cite{supple}. Thus, $t_{\langle ij\rangle}$ and $U$ are entirely determined by the Wannier wave functions of the lattice potential $V_\text{L}({\bf r})$, as well as the $s$-wave scattering length between the atoms \cite{supple}. For simplicity, we take the Gaussian approximation for the Wannier wave functions. $ i_x+ i_z$ being even or odd distinguishes two subsets of lattice sites in the $xz$ plane. The leading order effect of the cavity light is a relative shift of the potential energy between the even and the odd sites. This shift is described by the $\eta$-term in Eq.~(\ref{TB_Model}) where $\eta$ is the expectation value of $\eta_0\cos(k_0x)\cos(k_0z)$ with the Wannier wave function on a single site. The Hamiltonian for the entire system consists of both the atom part and the cavity field part, that is $\hat{\mathcal{H}}=\hat{\mathcal{H}}_{\text{BH}}-\delta_\text{c} \hat{a}^\dag\hat{a}$, where $\delta_\text{c}$ is the energy detuning between the cavity and the pumping lights. 

We apply the standard mean-field theory for the BHM part \cite{Sachdev} and for the cavity field, and find
\begin{align}\label{eq:MF}
&\frac{\hat{\cal H}_{\rm MF}}{N_\Lambda}=-(\varphi_\text{e} \hat{b}^\dag_\text{o}+\varphi_\text{o} \hat{b}_\text{e}^\dag+{\rm h.c.})-\mu_\text{e}\hat{n}_\text{e}-\mu_\text{o}\hat{n}_\text{o}\nonumber\\
&+\frac{U}{2}\hat{n}_\text{e}(\hat{n}_\text{e}-1)+\frac{U}{2}\hat{n}_\text{o}(\hat{n}_\text{o}-1)\nonumber\\
&+\frac{t_y(|\varphi_o|^2+|\varphi_e|^2)-(t_x+t_z)(\varphi_o^*\varphi_e+\varphi_e^*\varphi_o)}{2[t_y^2-(t_x+t_z)^2]}.
\end{align}
where the subscripts $o$ or $e$ denote for the odd or even site. The onsite energies are 
$\mu_\text{e}=\mu-\eta(\alpha+\alpha^*)$ and $\mu_\text{o}=\mu+\eta(\alpha+\alpha^*)$, 
and the mean value of the cavity field is $\alpha=\langle \hat{a}\rangle$. $2N_\Lambda$ is the total number of sites. 
The superfluid order parameters $\varphi_\text{e}$ and $\varphi_\text{o}$ on the even and odd sites respectively satisfy the mean-field equations 
\begin{align}
\varphi_e&=2(t_x+t_z)\langle \hat{b}_{\text{e}}\rangle+2t_y\langle \hat{b}_{\text{o}}\rangle,\label{eq:Selfconsis_SF1}\\
\varphi_o&=2(t_x+t_z)\langle \hat{b}_{\text{o}}\rangle+2t_y\langle \hat{b}_{\text{e}}\rangle.\label{eq:Selfconsis_SF2}
\end{align}
On the other hand, the dynamical cavity light field can leak out from the system with a decay rate $\kappa$, which makes the essential difference from the conventional BHM. In the regime of large decay rate $\kappa$, the cavity field obeys a steady equation $i\partial_t\alpha=\langle[\hat{a},\hat{\cal H}]\rangle-i\kappa\alpha=0$, which leads to 
\begin{equation}
\alpha=\frac{\eta N_\Lambda \left(\langle \hat{n}_{\text e}\rangle-\langle \hat{n}_{\text{o}}\rangle \right)}
{\delta_\text{c}+i\kappa}.\label{eq:SelfConsis_Alpha}
\end{equation}
It is clear that a non-zero $\alpha$ corresponds to a non-zero CDW order $\Theta\equiv\eta N_\Lambda \left(\langle \hat{n}_{\text e}\rangle-\langle \hat{n}_{\text{o}}\rangle \right)$.
Equations (\ref{eq:Selfconsis_SF1}), (\ref{eq:Selfconsis_SF2}) and (\ref{eq:SelfConsis_Alpha}) form a complete set of self-consistent equations, with which one can determine $\varphi\equiv\{\varphi_\text{e}, \varphi_\text{o}\}$ and $\alpha$. We can prove that, under the conditions that the parameters $t_{\langle ij\rangle}$ and $U$ of the Bose-Hubbard model are independent of $\alpha$, the state obtained from these set of self-consistent equations minimizes the total energy of the system $E=\langle \hat{\mathcal H}\rangle$ with a fixed atomic chemical potential $\mu$ \cite{supple}. 

\begin{figure}[t]\centering
\includegraphics[width=3.2in]
{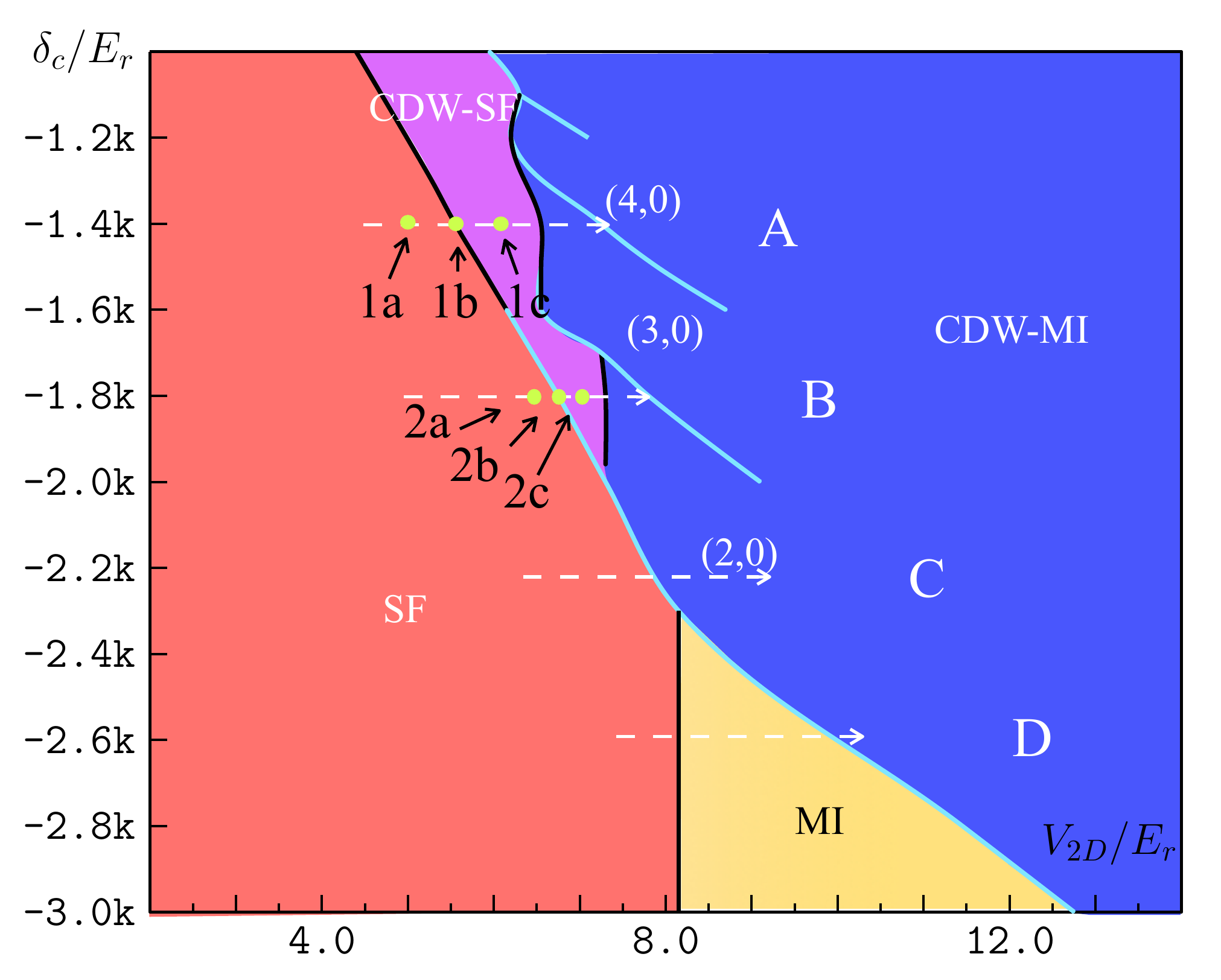}
\caption{A phase diagram between ``SF" (the red region), ``CDW-SF" (the purple region), ``MI" (the yellow region) and ``CDW-MI" (the blue region) as a function of $V_\text{2D}/E_\text{r}$ and $\delta/E_\text{r}$. The black lines denote the second order phase transitions while the blue thick lines denote the first order transitions. The symbol $(n_e,n_o)$ in the ``CDW-MI" phase denotes the atom number on the even and odd sites. Four dashed horizontal lines labelled by $A-D$ correspond to four different plots in Fig. \ref{Orderparameter} along which the order parameters are plotted. ``1a" to ``2c" mark the points where the energy curves are plotted in Fig. \ref{Energycurve} . Here we have taken $\mu=0.5U$, $V_y=20E_\text{r}$, $E_\text{r}=2\pi\times 4$kHz, $\kappa=2\pi\times 1.25$MHz, $V_c=0.0015E_r$ and $N_\Lambda=2\times 10^4$. \label{PhaseDiagram}}
\end{figure}

\textit{Phase Diagram:} In accord with the ETH experimental procedure \cite{ETH}, we fix the lattice depth along the $y$-direction $V_\text{y}$ (i.e. $t_y$ is fixed
), and increase the lattice potential $V_\text{2D}$ in the $xz$ plane. On one hand, as $V_\text{2D}$ increases, the hopping $t_x$ and $t_z$ decrease which drives the initially superfluiding system toward an insulating phase; and on the other hand, since the lattice beam along the $z$ direction also plays the role as the pumping beam, increasing $V_\text{2D}$ also drives the system toward the superradiant transition. The question is whether the two transitions occur separately or simultaneously. Before presenting the numerical results, we can first examine all scenarios by physical arguments. 

\begin{figure}[t]\centering
\includegraphics[width=2.5in]
{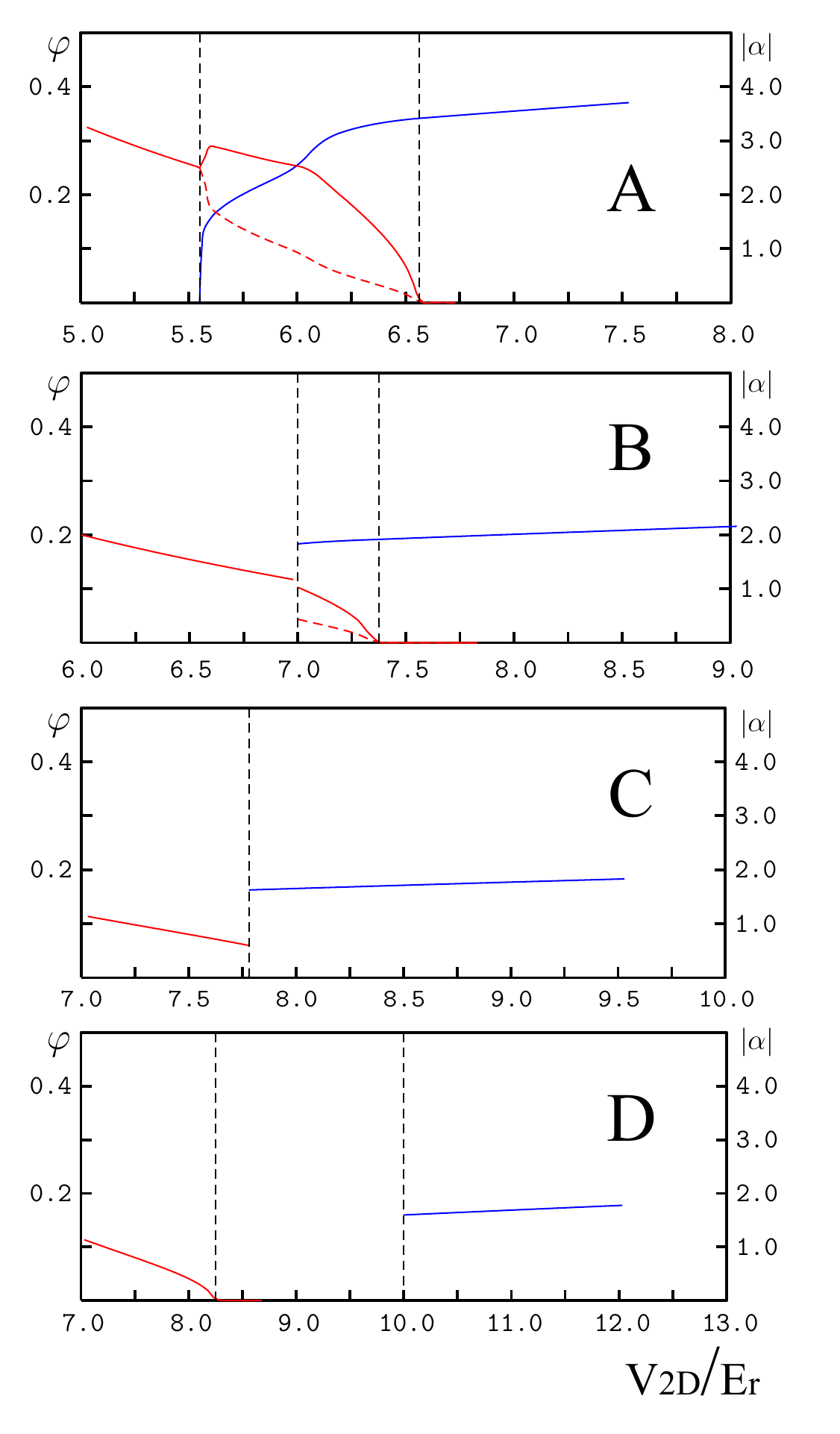}
\caption{Superfluid order parameters $\varphi_\text{e}$ and $\varphi_\text{o}$ (the red solid and the red dashed lines) and the cavity field mean value $\alpha$ as a function of $V_\text{2D}/E_\text{r}$ for four different $\delta/E_\text{r}$. (A-D) are plotted along the horizontal dashed lines labeled by $A-D$ in Fig. \ref{PhaseDiagram}. \label{Orderparameter}}
\end{figure}

a) If the two transitions happen simultaneously, two different sets of order parameters, i.e., $\alpha$ and 
$\varphi$, change at the same point which is most likely to render the transition a first order one. 

b) If the SF to MI transition happens before the superradiant transition, it will be a second order transition exactly the same as the one in the usual BHM. The successive superradiant transition will manifest itself as a MI to CDW-MI transition. Since in a MI phase the atom number on each site has to be integer, the number difference between the even and odd sites can only change discretely by an integer. Therefore, due to Eq.~(\ref{eq:SelfConsis_Alpha}), there must be a discontinuous change in $\alpha$ across the transition and the transition must be first order. 

c) If the superradiant transition happens first, whether the SF to CDW-SF transition is a first order one or a second order one is an open question. What we know from previous studies \cite{review,Tilman} is that in the absence of optical lattices, such a transition is second order. This conclusion indicates that if this superradiant transition happens far before the MI transition where the SF is still a weakly interacting one, the transition should still be second order. However, if the superradiant transition happens close to the MI transition, it is not known whether the strong correlation present in the superfluiding bosons can change the nature of this phase transition. Then, further increasing $V_\text{2D}$ will lead to a CDW-SF to CDW-MI transition. This is a SF to MI transition in the presence of a CDW order, and the order of this transition is also not clear.  

The main results of our numerical calculation is summarized by the phase diagram shown in Fig.~\ref{PhaseDiagram}. This phase diagram is obtained with fixed chemical potential $\mu$, and reveals the parameter regimes where different scenarios occur, and answers the questions raised in Scenario (c). 
In Fig.~\ref{Orderparameter} we plot both the superfluid order parameters $\varphi$ and the cavity field mean value $\alpha$ as functions of $V_\text{2D}$ for four different values of $\delta_\text{c}$.

In Fig.~\ref{Orderparameter}(a), $\delta_\text{c}$ is the smallest among the four. 
A small $\delta_\text{c}$ makes the superradiant transition easier. In this case, Scenario (c) applies, that is, the SF to CDW-SF transition first happens at a relatively small $V_\text{2D}$, and this transition is second order. Before the transition, $\varphi_\text{e}$ and $\varphi_\text{o}$ take the same value in the SF phase while they split after the transition. The second transition from CDW-SF to CDW-MI is also a second order one where the superfluid order parameter vanishes gradually, while in certain regime of $\delta_\text{c}$ it can also be first order. 

As $\delta$ increases, the SF to CDW-SF transition gradually moves to a larger $V_\text{2D}$ where correlations become stronger in the SF phase and suppress $\varphi$ to smaller values. In the case shown in Fig.~\ref{Orderparameter}(b), Scenario (c) still applies, but we find, surprisingly, that the transition becomes a first order one where both $\varphi$ and $\alpha$ display a jump. That is to say, the strong interaction effects in the SF phase make the superradiant transition first order. 

\begin{figure}[t]\centering
\includegraphics[width=3.5in]
{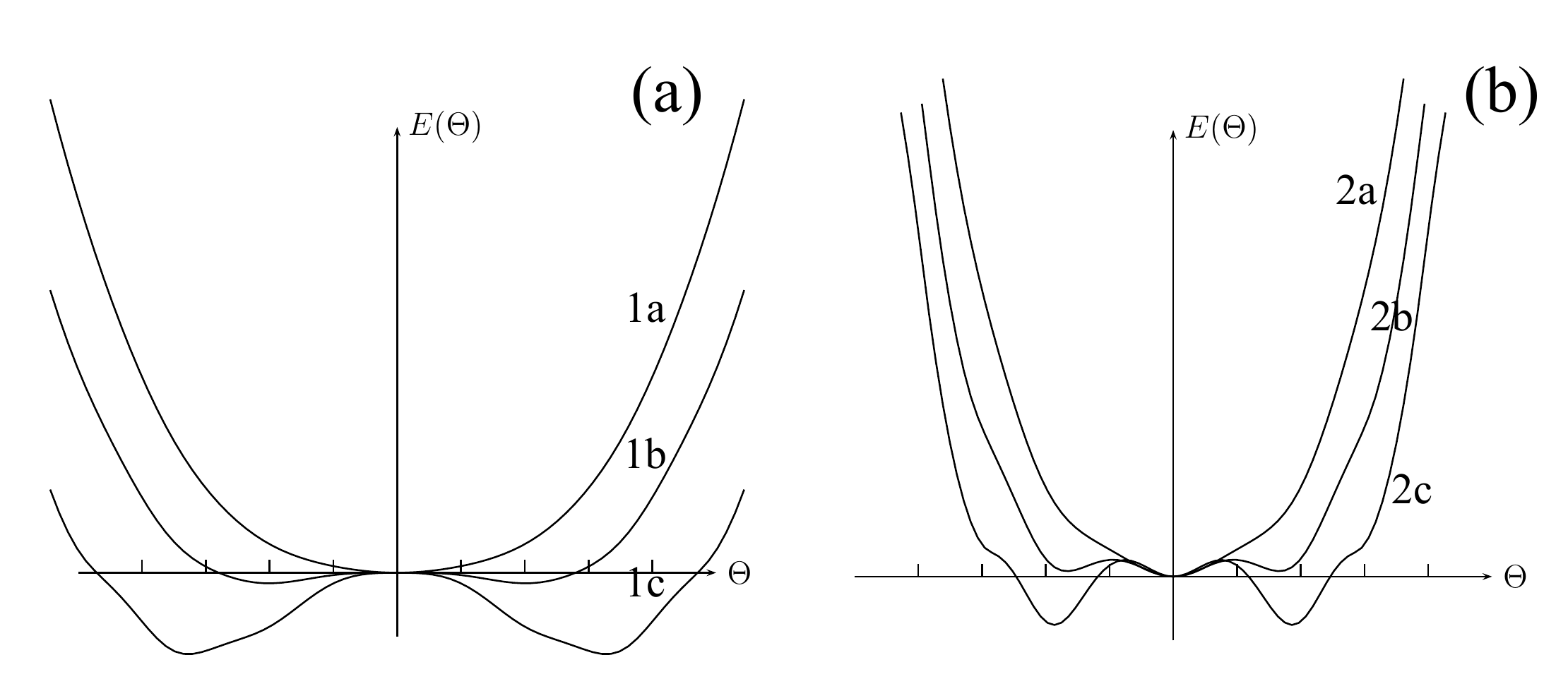}
\caption{The total energy $E=\langle \hat{H}\rangle$ as a function of CDW order parameter $\Theta$ on the vicinity of superradiance transition along Trajectory 1 and 2 in  Fig. \ref{PhaseDiagram}, respectively.  \label{Energycurve}}
\end{figure}

To manifest the difference between these two superrradiant transitions of different orders, in Fig.~\ref{Energycurve} we plot the energy curves $\mathcal{E}(\Theta)$  in the vicinity of the SF to CDW-SF transitions along Trajectory 1 and 2 in Fig. \ref{PhaseDiagram}, respectively. Figure \ref{Energycurve}(a) represents the typical behavior of $E(\Theta)$ when $\varphi$ is large. It is a standard Landau second-order transition with $E=a\Theta^2+b\Theta^4$, and $a$ changes sign with $b$ always being positive. Figure \ref{Energycurve}(b) represents the case when $\varphi$ is small. In this case $b$ changes sign first before $a$ changes sign, rendering a first order transition. To understand this difference, we recall that if $\varphi=0$, this transition becomes a MI to CDW-MI transition, and as we argued above, such a transition should be first order and therefore should have an energy curve similar to Fig.~\ref{Energycurve}(b). It turns out that such a behavior extends to finite but small $\varphi$, and therefore a SF with strong correlations in the vicinity of a MI transition also exhibits a first order behavior for the superradiant transition. In Fig.~\ref{PhaseDiagram}, the second order line and the first order line for the superradiant transition meet at an interesting tricritical point \cite{Book} where both $a$ and $b$ vanish. Figure \ref{Energycurve}(b) also shows that a substantial energy barrier for the first order transition persists after the transition point. The existence of such an energy barrier can be responsible for a wide coexistent region of MI and CDW-MI found in Ref.~\cite{ETH}.

When $\delta_\text{c}$ further increases, the SF to CDW-SF transition and the CDW-SF to CDW-MI transition merge together, after which Scenario (a) applies. As we show in Fig. \ref{Orderparameter}(c), $\varphi$ jumps from finite to zero while $\alpha$ jumps from zero to finite at a direct transition from SF to CDW-MI. 

Finally, at the largest value of $\delta_\text{c}$, the SF to MI transition takes place first, as we shown in Fig. \ref{Orderparameter}(d); Scenario (b) applies. 

\begin{figure}[t]\centering
\includegraphics[width=3.4in]
{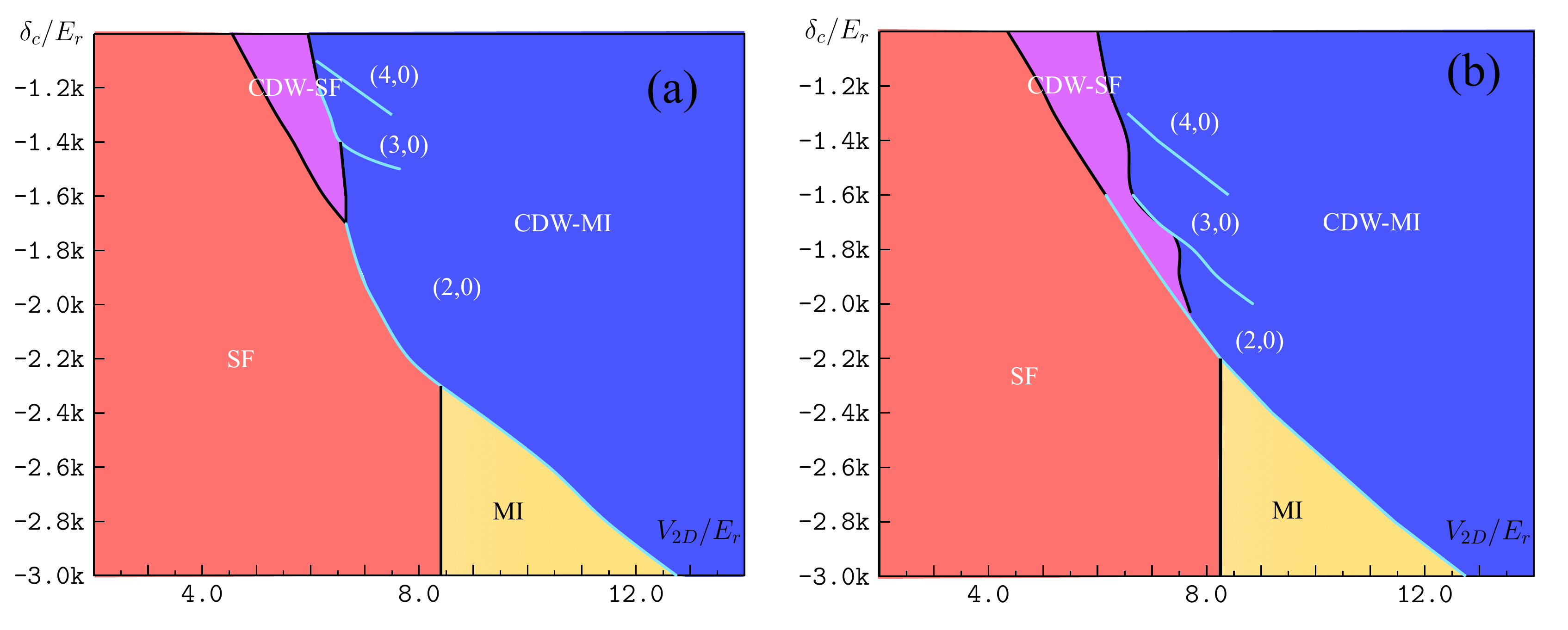}
\caption{Same phase diagram as Fig.~\ref{PhaseDiagram} for the same parameters, except for $\mu=0.25U$ (a) and $\mu=0.6U$(b)   \label{DifferentMu}.}
\end{figure}

In Fig.~\ref{DifferentMu} we show the same phase diagram but obtained for two different chemical potentials. We find that the basic structure of the phase diagram remains the same as Fig. \ref{PhaseDiagram}, and the difference is quantitatively. For instance, the regime of direct transition either expands, as in Fig. \ref{DifferentMu}(a), or shrinks, as in Fig. \ref{DifferentMu}(b).  The first order phase boundary between the SF to the CDW-SF phases can also vanish for certain range of chemical potential, such as in Fig. \ref{DifferentMu}(a). 

\textit{Final Remark.} We would like to point out that although our theoretical phase diagram is qualitatively similar to what is reported in Ref. \cite{ETH}, there are still some notable differences. First, in contrast to what is suggested in the experiment, in our theory, unless for a fine-tuned chemical potential, the four different phases generically shall not meet at the same point. Second, the first order SF to CDW-SF transition has not been observed yet. This difference may be due to the inhomogeneity of the experimental system in a harmonic trap, in which case the data are effectively an average over a range of chemical potentials, and also partly due to the finite resolution of the experiment measurements. Our theoretical predictions shall stimulate future experimental efforts in determining the phase diagram of this system more accurately. 

We should also mention the difference between the ETH setup and the Hamburg setup. In the Hamburg experiment, there is no external lattice beam along the cavity direction \cite{Hamburg}. That means, if not in the superradiant phase, there is no lattice potential along the cavity direction and hence the system can not be a MI. Therefore, there is no normal MI phase in their phase diagram. The CDW-SF to CDW-MI transition is driven by the cavity light. Therefore it happens when cavity field is sufficient strong and our approximation that $t$ and $U$ are independent of $\alpha$ can not apply. Finally, in the Hamburg setup $\kappa$ is relatively small, which makes both the Guassian approximation to the Wannier wave functions and the steady state approximation to the cavity field less accurate. Because of these considerable differences, theoretical study related to the Hamburg setup will be reported elsewhere.   

\textit{Acknowledgement.} We would like to thank Andreas Hemmerich for discussion. This work is supported by Tsinghua University Initiative Scientific Research Program, NSFC Grant No.~11325418 and No.~11474179.

\newpage
\begin{widetext}
\newpage

\section{Supplementary Materials}

\subsection{Tight-Binding Model Construction}

In this part of the supplementary materials, we are going to detail the construction of the tight-binding model, Eq.~(1), in the main text. The atoms in the cavity are subject to an optical potential $V_{\rm L}({\bf r})+V_{\rm C}({\bf r})$, where $V_{\rm L}({\bf r})=V_y\cos^2(k_0 y)+V_{2D}[\cos^2 (k_0x)+\cos^2 (k_0z)]$ is the external optical lattice potential and $V_{\rm C}({\bf r})=\eta_0\cos(k_0x)\cos(k_0z)(\alpha+\alpha^*)+V_c\cos^2 (k_0z) \alpha^*\alpha$ witih $\alpha=\langle \hat a\rangle$ being nonzero in the superradiant phase. The Hamiltonian of the system is given by
\begin{eqnarray}
\hat{\cal H}=\int d^3{\bf r}\hat{\Psi}^\dag({\bf r})\hat{H}_0\hat{\Psi}({\bf r})+g\int d^3{\bf r}\hat{\Psi}^\dag({\bf r})\hat{\Psi}^\dag({\bf r})\hat{\Psi}({\bf r})\hat{\Psi}({\bf r})-\delta_c \hat{a}^\dag \hat{a},\label{bareh}
\end{eqnarray}
where $\hat{\Psi}$ is the field operator for the atoms, and $\hat{H}_0={\bf p}^2/2m-\mu_0+V_{\rm L}({\bf r})+V_{\rm C}({\bf r})$. As in the deep lattice regime we concern, $V_{2D}$ and $V_y$ are much larger than $\eta_0(\alpha+\alpha^*)$ and $V_c \alpha^*\alpha$, we expand $\hat{\Psi}$ in terms of the lowest band of the optical potential $V_{\rm L}({\bf r})$, and approximate the corresponding Wannier wave function centering at the lattice site $\mathbf r_\mathbf j$ by a Gaussian form $\phi_\mathbf j(\mathbf r)=\Pi_{\ell=x,y,z}\Psi_{\ell}(r_{\ell}-r_{\mathbf j,\ell})$ and $\Psi_{\ell}(r_\ell)=(k_0^2\tilde\omega_{\ell}/\pi)^{1/4}e^{-\tilde\omega_{\ell} k_0^2r_\ell^2/2}$ with 
\begin{eqnarray}
&&\hspace{-3ex}\tilde\omega_{x}=\sqrt{|V_{2D}|/E_r},\hspace{3ex}\tilde\omega_{z}=\sqrt{|V_{2D}|/E_r},\hspace{3ex}\tilde\omega_y=\sqrt{|V_y|/E_r},
\end{eqnarray}
and $E_r=k_0^2/2m$. 

By substituting $\hat{\Psi}(\mathbf r)=\sum_j\hat{b}_j\phi_j(\mathbf r)$ into Eq.~(\ref{bareh}), we determine the parameters $\mu$, $\eta$, $t_{x,y,z}$ and $U$ of the tight-binding model, Eq.~(1), in the main text as follows. The shifted chemical potential is 
\begin{align}
\mu=\mu_0-\sum_{\ell=x,y,z}\tilde\omega_{\ell}E_r/2+|V_y|(1+e^{-1/\tilde\omega_{y}})/2+|V_{2D}|(1+e^{-1/\tilde\omega_{x}}),
\end{align}
where we have neglected the contribution $\sim V_c\alpha^*\alpha$ as well. And
\begin{align}
\eta=\eta_0e^{-1/4\tilde\omega_{x}-1/4\tilde\omega_{z}}.
\end{align}
The hopping amplitudes are given by
\begin{eqnarray}
t_{x}&=&-\int d^3{\bf r}\phi_{{\bf i}+{\bf e}_x}^*({\bf r})H_0\phi_{\bf i}({\bf r})
=\left[|V_{2D}|\left(\frac{\pi^2}4-e^{-1/\tilde\omega_x}\right)+\mu\right]e^{-\pi^2 \tilde\omega_x/4},
\end{eqnarray}
\begin{eqnarray}
t_{z}&=&-\int d^3{\bf r}\phi_{{\bf i}+{\bf e}_z}^*({\bf r})H_0\phi_{\bf i}({\bf r})
=\left[|V_{2D}|\left(\frac{\pi^2}{4}-e^{-1/\tilde\omega_z}\right)+\mu\right]e^{-\pi^2 \tilde\omega_z/4},
\end{eqnarray}
\begin{eqnarray}
t_{y}&=&-\int d^3{\bf r}\phi_{{\bf i}+{\bf e}_y}^*({\bf r})H_0\phi_{\bf i}({\bf r})
=\left[|V_{y}|\left(\frac{\pi^2}{4}-e^{-1/\tilde\omega_y}\right)+\mu\right]e^{-\pi^2\tilde\omega_y/4}.
\end{eqnarray}
In the regime $|V_y|$ is much bigger than $|V_{2D}|$, $t_y$ is much smaller than $t_{x,z}$, and the system is quasi-two dimensional. Finally, the onsite interaction energy is given as
\begin{eqnarray}
U=g\int d^3{\bf r}|\phi_{\bf i}({\bf r})|^2 |\phi_{\bf i}({\bf r})|^2=\bar{U}\sqrt{\tilde\omega_{x}\tilde\omega_{z}\tilde\omega_y},
\end{eqnarray}
with $\bar{U}=g k_0^3$. 

\begin{figure}[h]
\includegraphics[width=6in]
{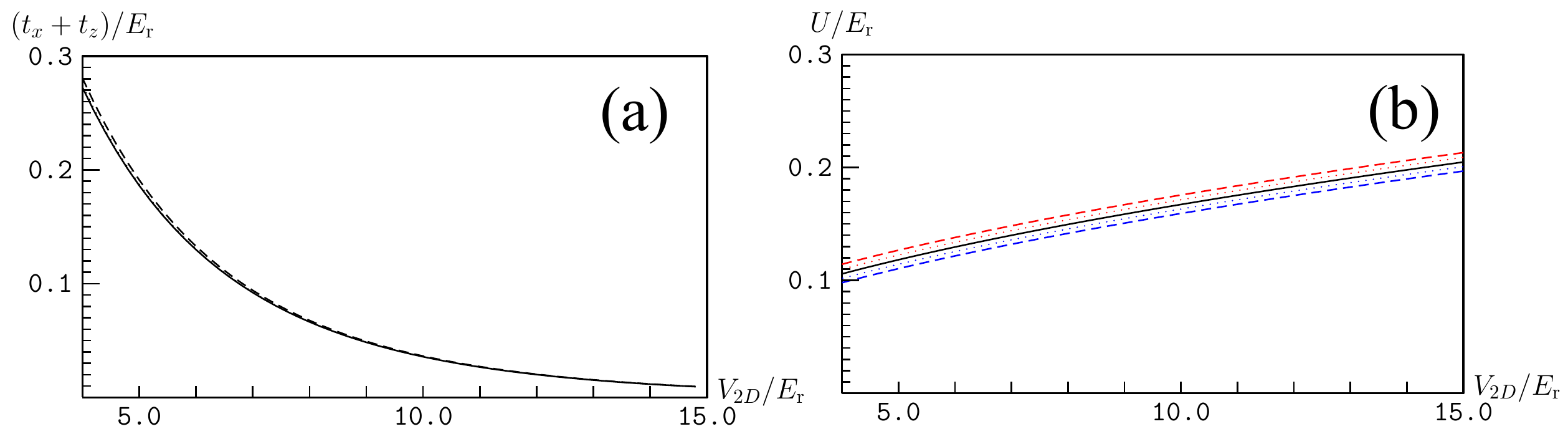}
\caption{(a) $t_x+t_z$ calculated for the lowest band of the optical potential $V_{\rm L}({\bf r})+V_{\rm C}({\bf r})$. The solid line is for $\alpha=0$, the dotted line for ${\rm Re}(\alpha)=4$, and the dashed line for ${\rm Re}(\alpha)=8$.
(b), $U$ calculated for the lowest band of the optical potential $V_{\rm L}({\bf r})+V_{\rm C}({\bf r})$. The solid line is for $\alpha=0$, the dotted line for ${\rm Re}(\alpha)=4$, and the dashed line for ${\rm Re}(\alpha)=8$. The two lines in red higher than the solid line are for the 
interaction energy on the even sites, the two lines in blue lower than the solid line are for the 
interaction energy on the odd sites.
}\label{just}
\end{figure}

Our approximation of neglecting the effects of the cavity field on the parameters $t_{x,y,z}$ and $U$ can be justified by comparing their values with the dependence on $\alpha$ taken into account. Figure \ref{just} shows that even with $|\alpha|$ is as big as $\sim 10$, the quantitative corrections to our approximation are within $10\%$. 
Since in the phase space of our interest a typical value of $|\alpha|$ is around $2$ to $4$, our approximation is in its reliable regime.
Furthermore, in the next part, we are going to show that by taking this approximation, the steady state solution satisfying Eqs.~(3) to (5) minimize the energy of the system. 
Therefore we apply this $\alpha$ independent parameter approximation through out our study. 

\subsection{Consistency for Matching Energy Minimum to Mean Field Self-consistent Solution}
In this section, we show explicitly how the approximation taking $t_{x,y,z}$ and $U$ to be $\alpha$ independent hopping makes the steady state solution and the requirement for the ground state being the energy minimal of the total energy curve meet together. In the superradiant phase, $\alpha=\langle \hat a\rangle$ is nonzero so is the density wave order parameter $\Theta=\eta N_\Lambda\langle(\hat n_\text{e}-\hat n_\text{o})\rangle$ of the atoms. 
Thus, we add the constraint $\Theta=\eta N_\Lambda\langle(\hat n_\text{e}-\hat n_\text{o})\rangle$, and need to minimize the total mean field hamiltonian
\begin{eqnarray}
{\cal\hat K}_{\rm MF}=-\delta_c \hat a^\dag \hat a+(\hat a^\dag+\hat a)\Theta+\lambda[\eta N_\Lambda(\hat{n}_\text{e}-\hat{n}_\text{o})-\Theta]+\hat S_0(\varphi, \mu,t,U,\hat b_{ i},\hat b_{ i}^\dag),
\end{eqnarray}
where $S_0$ is equal to ${\cal \hat H}_{\rm MF}$ given by Eq.~(2) in the main text except for $\alpha=0$, and $\lambda$ is a Lagrangian multiplier.
By requiring
\begin{align}
\partial \langle {\cal \hat K}_{\rm MF}\rangle/\partial\Theta=&\langle \partial_\Theta{\cal \hat K}_{\rm MF}\rangle=\alpha+\alpha^*-\lambda=0,
\end{align}
and using the steady state equation 
\begin{align}
i\partial_t\alpha=\langle[\hat{a},{\cal \hat K}_{\rm MF}]\rangle-i\kappa\alpha=-(\delta_c+i\kappa)\alpha+\Theta=0, 
\end{align}
we express the ground state energy $E=\langle {\cal \hat K}_{\rm MF}\rangle$ in terms of the order parameters $\varphi$ and $\Theta$ as
\begin{eqnarray}
E(\varphi,\Theta)=-\frac{\delta_c}{\delta_c^2+\kappa^2}\Theta^2+\frac{2\delta_c}{\delta_c^2+\kappa^2}\Theta\eta N_\Lambda(\langle \hat n_e- \hat n_o\rangle)+ \langle\hat S_0\rangle.
\end{eqnarray}
Since we take the approximation that $t_{x,y,z}$ and $U$ are independent of the cavity field, $\langle\hat S_0\rangle$ only depends on $\varphi$. The minimization $\partial_\varphi E(\varphi, \Theta)=\partial_\varphi \langle\hat S_0\rangle(\varphi)=0$ reproduces the mean field Eqs.~(3) and (4) in the main text. Most importantly, the other minimization 
\begin{eqnarray}
\partial_\Theta E(\varphi,\Theta)=-\frac{2\delta_c}{\delta_c^2+\kappa^2}[\Theta-\eta N_\Lambda (\langle \hat n_e- \hat n_o\rangle)]=0
\end{eqnarray}
is automatically satisfied due to the constraint $\Theta=\eta N_\Lambda\langle(\hat n_e-\hat n_o)\rangle$ implemented by introducing the 
Lagrangian multiplier $\lambda$.



On the other hand, if one includes the dependence of $t_{x,y,z}$ and $U$ on $\alpha$, there would be an extra term $\sim \partial_\Theta \langle\hat S_0\rangle$ contributing to $\partial_\Theta E(\varphi,\Theta)$, and generally the minimization of $E(\varphi,\Theta)$ with respect to $\Theta$ becomes inconsistent with the constraint $\Theta=\eta N_\Lambda\langle(\hat n_e-\hat n_o)\rangle$ unless $\kappa=0$.
This inconsistency lies in projecting the full Hamiltonian given by Eq.~(\ref{bareh}) to the lowest band. 
\end{widetext}


\begin{thebibliography}{99}

\bibitem{cavity}
H. Walther, B. T. H. Varcoe, B. G. Englert and T. Becker, Rep. Prog. Phys. {\bf 69}, 1325 (2006); and R. Miller, T. E. Northup, K. M. Birnbaum, A. Boca, A. D. Boozer and H. J. Kimble, J. Phys. B: At. Mol. Opt. Phys, {\bf 38}, S551 (2005)


\bibitem{Dicke}
R. H. Dicke, Phys. Rev. {\bf 93}, 99 (1954).

\bibitem{int_review}
I. Bloch, J. Dalibard, and W. Zwerger, Rev. Mod. Phys. {\bf 80}, 885 (2008).


\bibitem{Rit05} 
C. Maschler and H. Ritsch, Phys. Rev. Lett. {\bf 95}, 206401 (2005), C. Maschler, I. B. Mekhov, and H. Ritsch, Eur. Phys. J. D, {\bf 46}, 545 (2008).

\bibitem{Lew08}
J. Larson, B. Damski, G. Morigi and M. Lewenstein, Phys. Rev. Lett. {\bf 100}, 050401 (2008), J. Larson, S. Fernandez-Vidal, G. Morigi, M. Lewenstein, New Jour. Phys. {\bf 10}, 045002 (2008).

\bibitem{Simons09}
M. J. Bhaseen, M. Hohenadler, A. O. Silver, and B. D. Simons, Phys. Rev. Lett. {\bf 102}, 135301 (2009).

\bibitem{Simons10}
A. O. Silver, M. Hohenadler, M. J. Bhaseen, and B. D. Simons, Phys. Rev. A {\bf 81}, 023617 (2010).

\bibitem{Ciuti13}
A. L. Boit\'{e}, G. Orso, and C. Ciuti, Phys. Rev. Lett. {\bf 110}, 233601 (2013).

\bibitem{Hemmerich14}
M. R. Bakhtiari, A. Hemmerich, H. Ritsch, and M. Thorwart, Phys. Rev. Lett. {\bf 114}, 123601 (2014).

\bibitem{Benitez15}
S. F. Caballero-Benitez and I. B. Mekhov, Phys. Rev. Lett. {\bf 115}, 243604 (2015).

\bibitem{Chen}
Y. Chen, H. Zhai, and Z. Yu, Phys. Rev. A {\bf 91}, 021602(R) (2015).

\bibitem{Hamburg}
J. Klinder, H. Ke$\beta$ler, M. R. Bakhtiari, M. Thorwart, and A. Hemmerich, Phys. Rev. Lett. {\bf 115}, 230403 (2015).

\bibitem{ETH}
R. Landig, L. Hruby, N. Dogra, M. Landini, R. Mottl, T. Donner and T. Esslinger, arXiv: 1511.00007.

\bibitem{review}
H. Ritsch, P. Domokos, F. Brennecke, and T. Esslinger, Rev. Mod. Phys. {\bf 85}, 553 (2013).

\bibitem{Tilman}
F. Brennecker, T. Donner, S. Ritter, T. Bourdel, M. K\"{o}hl, and T. Esslinger, Nature (London) {\bf 450}, 268 (2007).

%

\bibitem{supple}
See supplementary material for (i) the parameters of BHM; and (ii) the proof that our self-consistent mean-field does minimize the total energy of this system.

\bibitem{Sachdev} S.~Sachdev, \emph{Quantum Phase Transitions}, 2nd ed.~(Cambridge University Press, Cambridge, 2011).



\bibitem{Book}
P. M. Chaikin and T. C. Lubensky, \emph{Principles of condensed matter physics} (Cambridge University Press, Cambridge, 1995).


\end{thebibliography}
\end{document}